# Estimating the Energy Requirements to Operate a Cryptanalytically Relevant Quantum Computer

Edward Parker and Michael J. D. Vermeer

## RAND Homeland Security Operational Analysis Center
An FFRDC operated by the RAND Corporation under contract with DHS

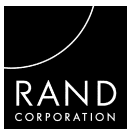

# Contents

# Figures and Tables

## Figures

## Tables

# Abstract

The academic literature contains many estimates of the resources required to operate a cryptanalytically relevant quantum computer (CRQC) in terms of rather abstract quantities like the number of qubits. But to our knowledge, there have not been any estimates of these requirements in terms of more familiar economic resources like money or electricity. We demonstrate that the electrical energy required to break one cryptographic public key can be decomposed into the product of two factors. There is an extensive literature of previous estimates for one factor, the spacetime volume, that range over about six orders of magnitude; we discuss some interesting patterns in these estimates. We could not find any quantitative estimates at all for the other factor, the average power consumption per qubit. By combining several data points from existing superconducting-transmon quantum computers and extrapolating them to enormously larger scales, we make an extremely rough estimate of a plausible value of about six watts/qubit consumed by an eventual superconducting-transmon CRQC. By combining this estimate with a plausible spacetime volume estimate of $5.9 \times 10^6$ qubit-days from the prior literature, we estimate that – even after expending the enormous costs to build a CRQC – running it would require about 125 MW of electrical power, and using it to break one public key would cost about \$64,000 for electricity alone at current prices. Even if a CRQC is eventually built, merely operating it would probably remain the domain of nation-states and large organizations for a significant period of time.

# Introduction

The use of a quantum algorithm known as *Shor's algorithm* to efficiently break existing public-key cryptography algorithms is one of the best-known eventual applications of quantum computers. Prototype quantum computers exist today, and they can already solve certain specialized problems in roughly the same amount of time as the world's most powerful classical supercomputers (although whether or not they can actually beat the best supercomputers is a matter of active debate among experts).[1] But these prototypes are still very far from being able threaten the security of modern public-key cryptography. An expert consensus study concluded that such a *cryptanalytically relevant quantum computer* (CRQC) is unlikely to become available before 2030,[2] and the U.S. National Security Agency has stated that it "does not know when or even if a … CRQC will exist."[3]

In this article, we provide some very rough estimates of the physical resources that might eventually be required for a CRQC to break public-key encryption of strengths that are currently considered sufficient to protect (for example) sensitive internet traffic. We believe that this topic is of interest to a broad range of audiences, including policy makers and quantum computing practitioners, so we have tried to keep this article fairly accessible to a wide range of audiences. We do not assume extensive technical background in either public-key cryptography or quantum computing, and we italicize technical terminology upon its first use.

For concreteness, we consider two public-key cryptography protocols that are in widespread use today and are considered secure against attacks from classical computers: the Rivest-Shamir-Adleman encryption algorithm using public keys that are 2048 bits long (*RSA-2048)* and elliptic-curve-cryptography (ECC), e.g. Diffie-Hellman key exchange, using public keys that are 256 bits long (*256-bit ECC*). For the purposes of this article, the reader only needs to know that RSA-2048 and 256-bit ECC are both widely used public-key cryptosystems that protect a large amount of highly sensitive internet traffic today, and both would be vulnerable to a CRQC implementing Shor's algorithm.

Many design challenges need to be solved before a CRQC can be built. The most basic challenge is simply scaling up the size of existing quantum computers. The largest quantum computer existing in March 2023 has 433 qubits[4] – the elementary hardware building blocks of a quantum processor – but the absolute bare minimum known number of qubits required to break (for example) RSA-2048 is 4,097.[5] Even scaling up to this number of qubits will require solving several engineering challenges. But in practice, many, many more qubits than this will be required. All existing quantum computer prototypes are known as *noisy intermediate-scale quantum* (NISQ) computers, and their quantum processors become corrupted by environmental noise within milliseconds of starting a computational run. Therefore, they are limited to extremely short (and therefore relatively simple) computations. But Shor's algorithm is an extremely complex algorithm that will require computations that run far longer than individual qubits can remain immune to environmental noise.

---

[1] Liu et al., "Validating quantum-supremacy experiments with exact and fast tensor network contraction."

[2] National Academies of Sciences, Engineering, and Medicine, *Quantum Computing: Progress and Prospects*.

[3] National Security Agency, "Quantum Computing and Post-Quantum Cryptography."

[4] IBM Quantum, "Quantum-centric supercomputing: The next wave of computing."

The solution to this problem is known as *quantum error correction* (QEC). There exist several known techniques (known as *error-correcting codes*) for networking together qubits in ways that allow for environmental errors to be detected and corrected before they corrupt the computation. Many of these techniques are well-understood theoretically, but quantum error correction is incredibly challenging to implement in practice, and only basic proofs of principle have yet been demonstrated experimentally.[6] QEC allows for *fault-tolerant* quantum computers, which use *logical qubits* that are extremely stable and resistant to environmental noise. In principle, logical qubits can be used to implement complex quantum algorithms like Shor's algorithm. But each logical qubit is composed of many *physical qubits* (the actual qubits that exist within the quantum computer). The exact number of physical qubits required to make one logical qubit depends on the hardware quality of the physical qubits as well as the error-correcting code. But as we will see, an algorithm as complex as Shor's algorithm will typically require well over one thousand physical qubits per logical qubit, so the need for QEC imposes an extremely high hardware overhead.

An additional complication is that several different basic physical architectures for qubits are being developed in parallel by different groups, such as superconducting-transmon qubits, trapped-ion qubits, neutral-atom qubits, photonic qubits, silicon-spin qubits, and topological qubits.[7] Each of these architectures have their own advantages and disadvantages along dimensions such as stability (i.e., resilience against environmental noise), speed of operation, ease of performing precisely controlled logic operations, feasible connectivity patterns, engineering practicality, and others, and they are each at a somewhat different level of technological maturity (although none of them are yet ready to be scaled up to cryptanalytically relevant sizes). It is therefore difficult to predict even the basic physical operating principles of a future CRQC, let alone details about its system architecture. Whichever qubit architecture proves the most feasible, building a CRQC will require an enormous amount of new fundamental systems engineering at every level – from the high-level algorithmic flow to the low-level engineering, networking, and control of individual qubits.[8]

---

[5] Gidney, "Factoring with n+2 clean qubits and n-1 dirty qubits." When we refer to the "size" of a quantum computer in this article, we always mean the number of qubits that it uses, not its physical size (although those two quantities scale roughly proportionately in certain regimes).

[6] Google Quantum AI, "Suppressing quantum errors by scaling a surface code logical qubit."

[7] Parker et al., "An Assessment of the U.S. and Chinese Industrial Bases in Quantum Technology."

It is therefore enormously challenging to make any estimates of the financial requirements to operate a future CRQC, even at the order-of-magnitude level of precision. As we will discuss below, we could not find any serious attempts to do so in the academic literature after a careful review. However, an estimate of the resources required to practically operate a CRQC has important implications for both research and development strategies and policy, especially with respect to efforts to address future cybersecurity risks. Therefore, even a framework for creating a highly uncertain estimate will be useful, provided that it can evolve over time as the field of quantum computing matures and uncertainties diminish. With those caveats, the rest of this section will attempt to give some extremely rough estimates for the economic costs of operating a future CRQC under a range of plausible assumptions. The final takeaway will not be a reliable point estimate, or even a precise range of uncertainty. Instead, we will demonstrate the huge range of resources that might be required under reasonable engineering assumptions.

## Intermediate Steps Toward a Cost Estimate

We did not attempt to estimate the (enormously high) upfront cost to build or purchase a CRQC[9] or the ongoing costs of maintenance, operator labor, administration, and other variables. However, we have estimated the marginal operating cost of the electricity required to break one cryptographic public key of realistic size. By far the biggest operating cost of modern high-performance-computing data centers is electrical power,[10] and as we will discuss, the same will probably be true of a CRQC. These simplifying assumptions are therefore probably reasonable.

One important caveat is that we also neglect the ongoing cost of physical inputs other than electricity, such as liquid helium. Some qubit architectures can operate at room temperature, but many of them can only operate at cryogenic temperatures that require liquid helium to maintain. A major global helium shortage began in 2022 that greatly increased helium prices, and the future price trajectory of helium is difficult to predict.[11] Also, starting around 2010, new *dry* or *cryogen-free* designs of dilution refrigerator became capable of recycling their liquid helium internally in a closed loop, greatly reducing their liquid helium consumption rate.[12] Similar improvements in refrigerator design could similarly decrease the required consumption rate of liquid helium, making the future helium requirements also difficult to predict. It is therefore possible that the ongoing operational cost of liquid helium, which we neglect in this article, could be comparable to the cost of electricity.

---

[8] Matsuura et al., "A systems perspective of quantum computing."

[9] Just *designing* an optimized quantum circuit to efficiently run Shor's algorithm, without even actually building or running it, already might require 1 GWh of energy for classical computation, which would cost $72,000 at current electricity prices. Paler and Basmadjian, "Energy Cost of Quantum Circuit Optimisation: Predicting That Optimising Shor's Algorithm Circuit Uses 1 GWh."

[10] Basmadjian, "Flexibility-based energy and demand management in data centers: A case study for cloud computing."

[11] Kornbluth, "Helium Shortage 4.0: What caused it and when will it end?"

[12] Nichols, "IBM's Goldeneye: Behind the scenes at the world's largest dilution refrigerator."

The energy consumed by a computational run of a quantum computer can be decomposed into the following product:

Energy consumed = runtime × (number of qubits) × (average power consumed per qubit). (1)

Here, the power per qubit includes not only the power directly consumed by each qubit to perform basic logic operations, but also all power consumption by hardware overhead, such as classical co-processing, cooling, and other forms of hardware systems control.

Equation (1) is always true by definition, but for a given hardware architecture, the average power consumed per qubit might vary as the system gets scaled to different sizes. In other words, the number of qubits and the average power consumed per qubit might not be independent variables. But implementation experts believe that for a given system architecture operating at a fault-tolerant scale, the power consumption requirements will probably scale roughly proportionately to the total number of qubits.[13] So we assume that the average power consumed per qubit will have a negligible dependence on the number of qubits, making equation (1) a useful decomposition. The average power per qubit could, however, depend strongly on the choice of qubit and system architecture.

The first two terms on the right-hand side of equation (1) (runtime and number of qubits) are often combined into a composite parameter called the *spacetime volume* of the quantum computer:[14]

Energy consumed = (spacetime volume) × (average power consumed per qubit).  (2)

The spacetime volume is a standard benchmark for summarizing the resource requirements for a quantum computer, and it has been extensively modeled. Moreover, there is a tradeoff between processor space (i.e., number of qubits) and time: for a given set of hardware performance parameters, a quantum circuit can be designed to operate more quickly at the expense of needing more qubits, or vice versa. As a rough rule of thumb, these tradeoffs tend to leave the spacetime volume approximately constant.[15]

---

[13] Hsu, "How Much Power Will Quantum Computing Need?"; Martin et al., "Energy Use in Quantum Data Centers: Scaling the Impact of Computer Architecture, Qubit Performance, Size, and Thermal Parameters."

[14] The spacetime volume is related to, but distinct from, the *quantum volume* metric that some companies use to quantify the performance of their quantum computers. Some authors use the term "time" to refer to the (dimensionless) abstract circuit depth, but in this article we mean the actual (dimensionful) runtime. Spacetime volume is often reported in units of megaqubit-days (millions of qubit-days), but in this paper we use units of qubit-days.

The rest of this chapter will attempt to roughly estimate plausible values for each of the factors in equation (2). As we will see, there is an extensive literature estimating the spacetime volume – although a very wide range of estimated values – but almost zero literature estimating the average power consumed per qubit.

## Estimating the Spacetime Volume Required for Breaking Public-Key Encryption

Around 2009, expert understanding of the resources required to execute Shor's algorithm on relevant public key sizes became accurate enough to begin making quantitative estimates of the required runtimes and number of physical qubits. Since then, there have been many estimates for the required runtimes and number of qubits for breaking modern public-key encryption. For example, Gidney and Ekerå (2021) give a detailed literature review of resource estimates for breaking RSA-2048, and Larasati and Kim (2021) give one for breaking elliptic-curve cryptography.

Table 1 gives a summary timeline of resource estimates for breaking RSA-2048. We see that the resource estimates have consistently dropped over time, and both the estimated qubit count and runtime fell by several orders of magnitude over 12 years. These estimates will probably continue to decrease as researchers continue to improve the implementation details, although it is difficult to estimate how fast.[16]

**Table 1. Timeline of Resource Estimates for Breaking One RSA-2048 Public Key**

| Source of Estimate | Number of Physical Qubits | Expected Runtime (days) | Spacetime Volume (qubit-days) |
|---|---|---|---|
| Van Meter et al. (2009) | $6.5 \times 10^9$ | 410 | $2.6 \times 10^{12}$ |
| Jones et al. (2010) | $6.2 \times 10^8$ | 10 | $6.2 \times 10^9$ |
| Fowler et al. (2012) | $1.0 \times 10^9$ | 1.1 | $1.1 \times 10^9$ |
| O'Gorman et al. (2017) | $2.3 \times 10^8$ | 3.7 | $8.5 \times 10^8$ |
| Gheorghiu et al. (2019) | $1.7 \times 10^8$ | 1 | $1.7 \times 10^8$ |
| Gidney and Ekera (2021) | $2.0 \times 10^7$ | 0.31 | $6.5 \times 10^6$ |

[15] Gheorghiu and Mosca, "A Resource Estimation Framework For Quantum Attacks Against Cryptographic Functions: Recent Developments."

[16] These sources made fairly similar – although not identical – assumptions about the hardware quality of the physical qubits; almost all of the reduction in resource requirements came from algorithmic and systems engineering improvements. See Appendix B of Gidney and Ekerå (2021) for a detailed technical discussion.

Table 2 summarizes many state-of-the-art resource estimates made between 2021 and 2023 for breaking either RSA-2048 or 256-bit elliptic-curve cryptography (ECC). Although 256-bit ECC offers somewhat higher security against classical computer attacks than RSA-2048, both are considered secure against classical attacks and both are in widespread use today.[17] We note that one expert assessed that Gidney's and Ekerå's estimate of 20 million qubits and eight hours was the best in the literature.[18]

The fourth column of Table 2 captures one important summary statistic for the assumed hardware quality of the physical qubits: the error rate for logic gate operations between two physical qubits (lower is better). As a benchmark for comparison, the superconducting-transmon, trapped-ion, neutral-atom, and silicon-spin architectures have each been experimentally demonstrated to achieve gate error rates between $10^{-3}$ and $10^{-2}$ for small numbers of qubits. No one has yet publicly demonstrated the ability to perform universal gate operations with well-characterized error rates using the photonic, superconducting-cat, or topological qubit architectures.[19]

Even though the basic engineering principles have been demonstrated for several different qubit architectures, building a CRQC using any of them will require enormous advances in qubit production and systems engineering. Some of the sources below assume completely new technologies that have not yet been demonstrated at any scale; these are indicated in the last column of Table 2. Specifically:

- Gouzien and Sangouard (2021) assume that the quantum processing unit is connected to a stable *quantum memory* that does not perform any logic operations but can store quantum states separately from the quantum processor, somewhat analogously to random-access memory (RAM) in a classical computer. They assume that the quantum memory contains 430 million storage modes with a two-hour lifetime. No such long-lived and high-capacity quantum memory has yet been demonstrated experimentally, but the authors propose a rare-earth-doped solid as a physical instantiation. Their estimated qubit count does not include any qubits that might be required to sustain the quantum memory.
- Some of Beverland et al.'s (2022) estimates assume topological qubits with extremely low error rates (see the notes below the table). Some of the building blocks of topological qubits, such as *Majorana zero modes*, have been demonstrated experimentally, but no actual topological qubits have been – although there are theoretical reasons to believe that if they can be built, then they will experience very low error rates.[20]

---

[17] IBM, "Size considerations for public and private keys."

[18] Goodin, "RSA's demise from quantum attacks is very much exaggerated, expert says."

[19] Parker et al., "An Assessment of the U.S. and Chinese Industrial Bases in Quantum Technology."

- Gouzien et al. (2023) assume another qubit hardware architecture, known as "cat qubits," which consist of superconducting waveguide microwave resonators or optical cavity modes that are coupled together nonlinearly using Josephson junctions. These have not yet been experimentally demonstrated to be capable of performing well-characterized logic operations.

---

**Table 2. Summary of Recent Resource Estimates for Breaking One RSA-2048 or 256-Bit ECC Public Key**

| Source of Estimate | Encryption Algorithm Cracked | Qubit Hardware Architecture[a] | Physical Two-Qubit Error Rate | Number of Physical Qubits | Runtime (days)[b] | Spacetime volume (qubit-days) | Non-demonstrated Technology? |
|---|---|---|---|---|---|---|---|
| Gheorghiu and Mosca (2021) | RSA-2048 | Superconducting | $10^{-3}$ | $2.2 \times 10^7$ | 0.014 | $3.1 \times 10^5$ | - |
| | | | $10^{-5}$ | $8.7 \times 10^6$ | 0.0063 | $5.5 \times 10^4$ | |
| Gidney and Ekerå (2021) | RSA-2048 | Superconducting | $10^{-3}$ | $2.0 \times 10^7$ | 0.30 | $5.9 \times 10^6$ | **-** |
| Gouzien and Sangouard (2021) | RSA-2048 | Superconducting | $10^{-3}$ | $1.3 \times 10^4$ | 177 | $2.3 \times 10^6$ | Separate stable quantum memory |
| Webber et al. (2022) | RSA-2048 | Trapped ion | $10^{-3}$ | $6.5 \times 10^8$ | 10 | $6.5 \times 10^9$ | - |
| | 256-bit ECC | Superconducting | $10^{-3}$ | $1.9 \times 10^9$ | 0.007 | $1.3 \times 10^7$ | |
| | | | | $3.2 \times 10^8$ | 0.042 | $1.3 \times 10^7$ | |
| | | | | $1.3 \times 10^7$ | 1 | $1.3 \times 10^7$ | |
| | | | $10^{-4}$ | $3.3 \times 10^7$ | 0.042 | $1.4 \times 10^6$ | |
| Beverland et al. (2022)[c] | RSA-2048 | Trapped ion | $10^{-3}$ | $3.7 \times 10^7$ | 2260 | $8.4 \times 10^{10}$ | - |
| | | | $10^{-4}$ | $8.6 \times 10^6$ | 1100 | $9.5 \times 10^9$ | |
| | | Superconducting or spin | $10^{-3}$ | $3.7 \times 10^7$ | 1.5 | $5.6 \times 10^7$ | |
| | | | $10^{-4}$ | $8.7 \times 10^6$ | 0.75 | $6.5 \times 10^6$ | |
| | | Topological | $10^{-4\,(d)}$ | $2.6 \times 10^7$ | 0.625 | $1.6 \times 10^7$ | Topological qubits |
| | | | $10^{-6\,(d)}$ | $6.2 \times 10^6$ | 0.3 | $1.9 \times 10^6$ | |
| Gouzien et al. (2023)[c] | 256-bit ECC | Cat | $10^{-5\,(e)}$ | $1.3 \times 10^5$ | 0.375 | $4.9 \times 10^4$ | Cat qubits |
| | RSA-2048 | | | $3.5 \times 10^5$ | 4 | $1.4 \times 10^6$ | |

| | SOURCE: Adapted from Table 2 of Gidney and Ekerå (2021). Complete citations for individual rows in this table are given in Gidney and Ekerå. |
| --- | --- |
| | NOTES: [a] "Superconducting" refers to superconducting-transmon qubits; superconducting cat qubits are referred to as "cat". The only resource estimate for neutral-cold-atom qubits that we found in the literature was in Suchara et al. (2013), but they only considered attacking RSA-1024, and we judged that their methodology was too out of date to be comparable to the more recent assessments reported here.<br>[b] Some of these sources assume that only a single run of Shor's algorithm is required, but others incorporate the possibility that the first run fails (due to intrinsic quantum indeterminacy, not hardware errors) and so multiple runs are required. This possibility increases the expected runtime by at most 50%.<br>[c] Beverland et al. (2022) and Gouzien et al. (2023) are still preprints that have not yet been peer-reviewed.<br>[d] These two-qubit gate error rates correspond to the Clifford gates, which are topologically protected and so experience very few errors. The non-Clifford gates, which are not topologically protected, are assumed to experience error rates of 5% or 1% in the two cases. These much higher error rates for the non-Clifford gates remove much of the benefit of the topological protection of the Clifford gates.<br>[e] The actual two-qubit gate error rate is actually $8.4 \times 10^{-3}$. But for cat qubits, the more relevant error rate is the ratio of the single-photon and the double-photon loss rates, which is assumed to be $10^{-5}$. We report this latter error rate in the table because it is more directly comparable to the error rates for the other entries. (E.g., it enters into the formula for the logical error rate under the repetition code in the same way that the two-gate error rate enters into the formula for the superconducting qubits' logical error rate under the surface code.) |

These sources also contain insights into the costs of other resources beyond qubits and time. For example, Webber et al. (2022) note that factoring RSA-2048 with a trapped-ion quantum computer with a two-qubit gate error rate of $10^{-3}$ would require atom traps that would take up 3600 m$^2$ of tabletop space, or about half the size of a soccer field. If the two-qubit gate error rate is lowered to $10^{-4}$, then the ion traps would take up "only" 324 m$^2$ of tabletop, or about the size of a doubles tennis court. The authors estimate that with a few new technology developments such as mid-range qubit connectivity, the required area could be reduced to 6.25 m$^2$, or the area of two office desks. They did not provide a qubit estimate for this case, but if we assume a constant number of qubits per unit area of ion traps, then this would extrapolate to about $5.9 \times 10^6$ physical qubits.

We can organize the estimates in Table 2 into three broad categories that represent three levels of ambition regarding technical hardware improvement beyond today's levels. All three of these scenarios are ambitious, given the enormous systems engineering requirements for a CRQC.

1. Ambitious scenarios: Physical qubits and logic gate operations improve to reach a two-qubit logic gate error rate of $10^{-3}$, which is within one order of magnitude of currently demonstrated error rates.

2. More ambitious scenarios: The two-qubit logic gate error rate reaches $10^{-4}$, which is more than one order of magnitude better than currently demonstrated error rates.

3. Most ambitious scenarios: (a) the two-qubit logic gate error rate reaches $10^{-5}$, and/or (b) fundamentally new technologies that have not yet been demonstrated experimentally (such as the three listed above) become feasible.

Figure 1 summarizes the estimates reported in Table 2. The axes represent the runtime to break one public key and the number of physical qubits required on log-log scales. The dashed lines represent contours of constant spacetime volume. The marker colors indicate the technology improvement assumptions, and the marker shapes indicate the qubit architecture.

**Figure 1. Summary of Resource Estimates to Break One RSA-2048 or 256-Bit ECC Public Key**

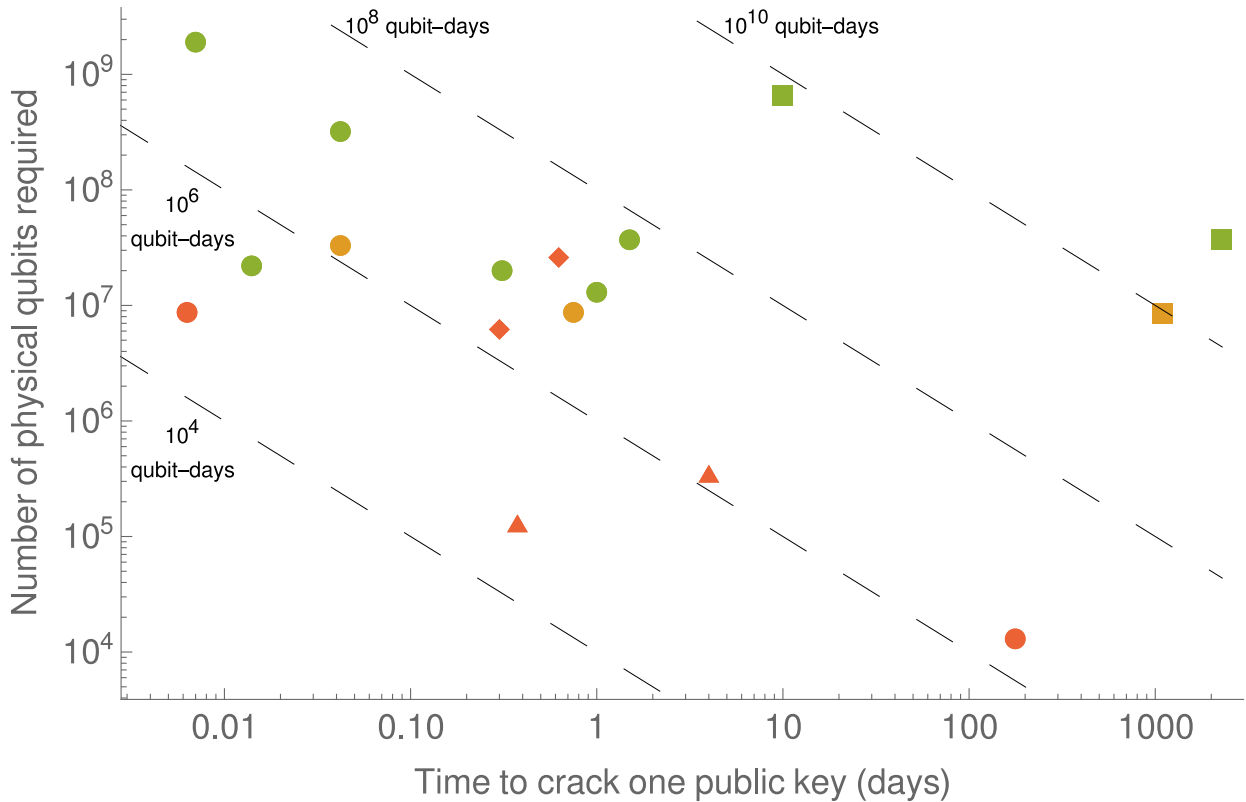

SOURCE: RAND analysis of sources listed in Table 2.
NOTES: Marker colors indicate the three technology advancement scenarios described in the main text: ambitious scenarios (green), more ambitious scenarios (yellow), or most ambitious scenarios (red). Market shapes indicate the qubit architectures: superconducting-transmon (circles), trapped-ion (squares), topological (diamonds), or cat (triangles). Dashed lines indicate contours of constant spacetime volume.

As a benchmark for comparison: as mentioned above, the largest quantum computer existing in March 2023, IBM's Osprey processor, has 433 qubits.[21] The runtime of a NISQ computer like Osprey is not very well-defined, since there is some subjectivity involved in judging when the noise has degraded the quantum state beyond the point of usefulness. But if we use the qubit coherence time as a rough upper bound for the runtime, then Osprey can run for about 75 μs, for a roughly equivalent spacetime volume of $4 \times 10^{-7}$ qubit-days.[22] We make the important caveat that NISQ devices are not directly comparable to the fault-tolerant quantum computers in Table 2. But if we were to plot the Osprey processor in Figure 1, then it would be well below the bottom of the chart – and so far to the left of the current vertical axis that including the Osprey processor would require us to expand the chart to nearly three times its current width, even using a logarithmic axis scale.

---

[21] IBM Quantum, "Quantum-centric supercomputing: The next wave of computing."

[22] Other IBM quantum processors have fewer qubits but longer coherence times, for an approximately similar spacetime volume. Hu, "IBM's biggest quantum chip yet could help solve the trickiest math problems."

The Figure has a few clear takeaways:

- The estimated runtimes required to crack one public key span many orders of magnitude, from 10 minutes to six years.
- The estimated required qubit counts also span many orders of magnitude, from $10^5$ to $10^9$ qubits.
- Almost all the green data points – which assume physical qubit hardware quality near currently demonstrated capabilities – require a spacetime volume above $10^6$ qubit-days.[23]
- Most of the estimates that assume superconducting qubits that are close to currently demonstrated hardware quality (green circles) require a spacetime volume of approximately $10^7$ qubit-days – about 14 orders of magnitude beyond our current capabilities. More optimistic assumptions for advances in hardware quality (the yellow circles) require between $10^6$ and $10^7$ qubit-days.
- Trapped-ion quantum computers would require a roughly comparable number of qubits as superconducting-transmon quantum computers but would require $10^{10}$ qubit-days or more, because they would require much longer runtimes due to their much slower logic operations.
- Even if topological qubits can be successfully developed, they may still require more hardware resources than superconducting-transmon qubits do, unless the error rates for the topological qubits' non-Clifford gates, which are not topologically protected against noise, can be brought well below $10^{-2}$.

## Estimating the Required Power Per Qubit in a Cryptanalytically Relevant Quantum Computer

The average power consumed per qubit in a CRQC is extremely difficult to estimate without a much better understanding of both the individual qubit architecture and the full system design, including classical hardware and software control aspects such as input/output, logic gate programming, and quantum error correction. Trying to estimate the directly physical resource requirements (such as power consumption) for a CRQC before we know the optimal qubit architecture is somewhat like trying to estimate the resource requirements for a classical supercomputer without knowing whether its basic logic operations will be physically implemented by electronic transistors, vacuum tubes, or mechanical gears. As such, we were able to find very little academic literature making serious quantitative estimates of the power consumption for future quantum computers.

---

[23] One green data point, corresponding to the first row of Table 2 (Gheorghiu and Mosca, 2021), lies below the $10^6$ qubit-days contour. This is primarily because the authors assumed a very fast error-correction cycle time of only 200 ns. The other authors all assumed a slower cycle time of 1 μs. Adjusting Gheorgiu and Mosca's estimate to use the same cycle time as the other sources would multiply their runtime by a factor of 5 and bring it up to 1.7 hours, which would bring their spacetime volume estimate slightly above the $10^6$ qubit-days contour – still significantly more optimistic than the other green markers, but in much closer agreement than in Figure 1.

Villalonga et al. (2020) performed a careful comparative benchmarking of the electrical power consumed when the same mathematical problem was solved by Google's Sycamore NISQ computer and by IBM's Summit supercomputer at Oak Ridge National Laboratory, which at the time was the fastest classical supercomputer in the world. They found that for one problem instance, Summit consumed an average of 8.65 MW of power over 2.44 hours, for a total energy consumption of 21.1 MWh, while Sycamore consumed an average of 15 kW over 1 minute and 41 seconds, for a total energy consumption of 420 Wh. Sycamore therefore beat Summit's performance by a factor of 578 for power consumption, 87 for runtime, and a factor of $578 \times 87 \approx 50200$ for energy consumption. But Villalonga et al.'s (2020) analysis did not cover future quantum computers, which will have very different system architectures.

Jaschke and Montangero (2022) perform a more general parametric analytical analysis of quantum computers' power consumption for several different qubit architectures, but their analysis still only applies to NISQ computers, and their assumptions fix the total power consumption to that of existing NISQ systems, so their analysis does not apply to CRQCs.

The most comprehensive analysis of the power requirements for a large quantum computer that we found was by Martin et al. (2022), who performed an in-depth analysis of the distribution of power usage across elements such as direct computation and system cooling for several qubit architectures, starting from basic thermodynamic principles. One of Martin et al.'s (2022) main findings was that fault-tolerant quantum computers will probably use more energy for cooling than for direct computation, although the exact ratio is extremely sensitive to the system architecture. The opposite is true for existing high-performance computing data centers, whose power consumption for cooling is only 10-30% of the power consumption for the electronics. But unfortunately, they state that "The required power for electronics per qubit $q$ that will be seen in future quantum systems is challenging to estimate … the value of $q$ for future quantum systems cannot be reliably determined."[24] They therefore do not attempt to estimate the absolute power consumption for the system, but only the relative distribution of power consumption across various subfunctions. The main empirical quantitative inputs that they use are the expected operating temperatures of various qubit architectures: 10-20 mK for superconducting qubits and 4 K for trapped-ion qubits.

---

[24] Note that Martin et al.'s variable $q$ is not the same as the "average power consumed per qubit" variable in our equations (1) and (2). Their $q$ only incorporates the power *for electronics* that is directly used to power the qubit's logic operations, while our power includes all overhead power consumption (which will likely be primarily used for cooling). The equations in Martin et al.'s article can be used to approximately convert between these two notions of power consumption.

With so little concrete data available, we are forced to be highly speculative and to extrapolate our very few empirical data points far beyond the regimes where they are likely to hold. For the remainder of this section, we will only consider superconducting-transmon quantum computers, since they are currently the qubit architecture with the best-documented performance characteristics.

During its "quantum supremacy" run in 2019, Google's Sycamore quantum computer consumed an average of about 26 kW of electrical power, of which about 10 kW went directly to the dilution refrigerator's mechanical compressor, 10-13 kW went to cooling the chilled water for the refrigerator's compressor and pumps, and 3 kW went to the supporting classical electronics.[25] D-Wave's superconducting quantum annealers are not universal quantum computers, but they operate at similar temperatures and use roughly similar dilution refrigerator technology, and the D-Wave Advantage consumes a very similar 25 kW of power as Sycamore.[26]

The power consumption of a given dilution refrigeration is essentially independent of the number of qubits operating inside of it.[27] Indeed, Sycamore contained 53 functional qubits while the D-Wave Advantage contains over 5000 qubits, but they consume about the same amount of power. But there is a limit to how many qubits can fit inside one dilution refrigerator, so a large superconducting-transmon quantum computer will require either an *extremely* large dilution refrigerator, or many refrigerators networked together via quantum channels that are capable of entangling together qubits in different refrigerators, which is yet another piece of systems architecture that needs to be developed before a CRQC becomes feasible.

IBM's project Goldeneye is developing much larger dilution refrigerators that will eventually hold up to one million qubits, and IBM completed an initial proof-of-concept test of a smaller version in September 2022.[28] But given the estimates in Table 2, even if such a huge dilution refrigerator could be engineered, then it would probably still not be big enough to contain a CRQC. So, a CRQC built out of superconducting qubits would probably require networking together multiple dilution refrigerators. We can very roughly estimate the power consumption of such a system by separately estimating (a) the required number of dilution refrigerators and (b) the power draw of each dilution refrigerator, and then multiplying those two estimates together.

The largest superconducting-transmon quantum computer – or indeed any type of quantum computer – existing in March 2023 is IBM's Osprey computer, which contains 433 qubits in one dilution refrigerator. This is probably not yet reaching the physical space limits of a feasible dilution refrigerator; in 2019, Krinner et al. estimated that existing dilution refrigerators could be modified to fit about one thousand superconducting qubits with reasonably achievable engineering improvements. Moreover, IBM has also previewed such a concept for networking together modular dilution refrigerators known as IBM Quantum System Twos.[29] In this system, each modular dilution refrigerator would contain up to 4,158 qubits, and up to three System Twos' processors could be coupled together to achieve well over 10,000 qubits. IBM has announced that they will unveil their Quantum System Two in 2023.

A superconducting-qubit CRQC with 20 million qubits will certainly not look anything like 4,810 IBM Quantum System Twos networked together. Fundamentally new networking and control paradigms, including quantum error correction, will need to be developed before a CRQC becomes feasible. Nevertheless, the IBM Quantum System Two is set to be announced this year (2023) and so is probably achievable with current technology, and we can use it as an *extremely* rough benchmark to estimate the power requirements for cooling large-scale superconducting-qubit quantum computers with foreseeable technology. If 4,158 qubits will be able to fit into each IBM Quantum System Two dilution refrigerator, and we assume that each refrigerator will draw power comparable to Google's and D-Wave's dilution refrigerators' consumption of about 26 kW, then we get a power consumption of 6.25 watts/qubit – about the power consumption of one LED light bulb.[30]

This estimate incorporates all classical control and cooling power overhead except for the additional overhead for quantum error correction, which no current NISQ device uses. Researchers have demonstrated preliminary steps toward QEC, but none of these demonstrations have yet significantly extended the qubits' coherence times, so it is very difficult to estimate the additional power overhead required for the classical processing side of large-scale QEC.[31] The only discussion that we found for these power requirements was in Beverland et al. (2022): "For a system with a few million qubits, we estimate that several terabytes per second of bandwidth will be required between the quantum and classical planes. Furthermore, processing these measurements at a rate that is sufficient to effectively correct errors demands petascale classical computing resources that are tightly integrated with the quantum machine." Petascale operating speeds would place such a control system within about the top 500 supercomputers existing today, so the overhead from classical processing for error correction might make a significant contribution to a CRQC's total power consumption. On the other hand, improved qubit and refrigeration design could well shrink the power consumption per qubit far below current levels.

---

[29] IBM Research, "IBM Quantum System Two."

[30] Of course, if IBM's Goldeneye program or any other group succeeds in expanding dilution refrigerators to be able to fit in tens of thousands or even a million qubits, then there will presumably be economies of scale that drive the power per qubit well below current levels. But we have no way to estimate the power draw of a hypothetical one-million-qubit dilution refrigerator.

[31] Google Quantum AI, "Suppressing quantum errors by scaling a surface code logical qubit."

So, it is difficult to tell whether our rough estimate of 6.25 watts/qubit is more likely to be an overestimate or an underestimate.

For concreteness, we will use Gidney's and Ekerå's estimate of 20 million qubits and 7.1 hours as a benchmark for breaking one public key, since (as mentioned above) one expert assessed that to be the best estimate in the literature. If we simply extrapolate our estimate of 6.25 watts/qubit out linearly, we get that a superconducting-qubit CRQC would require 125 MW of electrical power. This is nearly 10 times the power consumed by the Summit supercomputer, about the power consumption of a Boeing 747 aircraft in flight, and about a quarter of the power produced by a typical coal-fired power plant. Over a 7.1-hour expected runtime, such a computer would consume 890 MWh of electrical energy. At the 2022 average U.S. industrial price of electricity of 7.19¢/kWh,[32] breaking one public key would cost $64,000 in electricity alone.

## Discussion

We have demonstrated quantitatively that there is enormous uncertainty in both of the factors that go into equation (2) to estimate the electrical energy required to use a CRQC to break one public key. The first factor, the spacetime volume, has an extensive literature of previous estimates that range over about six orders of magnitude, from $10^5$ to $10^{11}$ qubit-days. Moreover, history suggests that these estimates could change (probably downward) by orders of magnitude as our implementation designs improve. For the second factor, the average required power per qubit, we could not find any previous estimates at all. By combining information from multiple sources – including systems that have not yet actually been demonstrated – and extrapolating current power consumption values *far* beyond the scale of existing systems, we made an *extremely* rough estimate of 6.25 watts/qubit for superconducting-transmon quantum computers. This estimate completely neglected both the unknown economies of scale from producing larger systems (which would presumably push the value down) and the unknown additional classical processing overhead for quantum error correction (which would presumably push the value up). Combining this estimate with a plausible spacetime volume estimate of $5.9 \times 10^6$ qubit-days for a superconducting-transmon computer yields an electrical energy estimate of about 890 MWh per key broken, which costs about $64,000 at today's electricity prices. To our knowledge, this is the first quantitative estimate of this energy cost reported in the academic literature.

It is entirely reasonable to question the utility of this final estimate, given the number of separate assumptions, approximations, and extrapolations that went into it – some of which have orders of magnitude of uncertainty. We nevertheless believe that it is important to try to make resource estimates in practical terms that can be clearly understood by diverse audiences with varying levels of technical expertise in computing. Long-term strategies and policies are currently being crafted to plan for a future that includes CRQCs,[33] and the practical requirements for operating a CRQC provide important context for that planning, especially with respect to addressing cybersecurity risks. Significant literature exists that estimates resource requirements in terms of qubits, runtimes, and error rates, but many audiences will find it challenging to engage with the concept on those terms. Estimates of power usage and cost could communicate the practical implications of a CRQC more effectively than estimates of spacetime volume. Whether or not our final estimate is certain enough to be useful for all audiences, our analysis can serve as a useful framework for assessing these practical requirements. It can and should evolve over time as technology matures and further analysis allows us to determine the factors with greater clarity.

---

[32] U.S. Energy Information Administration, "Monthly Energy Review February 2023."

To those ends, we believe we can take away two messages with confidence:

First, there is enormous and qualitative uncertainty as to the level of resources that will be required to operate a CRQC.

Second, even if a CRQC does become feasible and gets built, then it will probably require very significant ongoing operating costs. There will still be many more technical advances required before breaking public-key cryptography becomes truly cheap (if it ever reaches that point at all). For the foreseeable future, we are extremely unlikely to reach a point where individual actors can break public-key cryptography using their own quantum computer. If a CRQC does get built, then there will probably be a significant period of time afterward when operating it will remain the domain of nation-states or similarly well-funded organizations.[34]

## Acknowledgments

The authors thank Nicolas Robles and Patricia Stapleton for helpful feedback during the drafting of this article. This research was sponsored by the National Risk Management Center within the Cybersecurity and Infrastructure Security Agency of the United States Department of Homeland Security, and it was conducted through the Infrastructure, Immigration, and Security Operations Program of the RAND Homeland Security Research Division.

---

[33] The White House, "National Security Memorandum on Promoting United States Leadership in Quantum Computing While Mitigating Risks to Vulnerable Cryptographic Systems," May 4, 2022.

[34] As a point of comparison, the 2012 Flame malware used a computationally intensive hash collision that was estimated to require $200,000 in computing power to exploit. From this fact, one cybersecurity analyst concluded that "There is little doubt that Flame was created by a nation state with considerable technical resources." Goodin, "Flame's crypto attack may have needed $200,000 worth of compute power"; Stiennon, "Flame's MD5 collision is the most worrisome security discovery of 2012."

# Abbreviations

| | |
|---|---|
| CRQC | cryptanalytically relevant quantum computer |
| ECC | elliptic-curve cryptography |
| NISQ | noisy intermediate-scale quantum [computer] |
| QEC | quantum error correction |
| RSA | Rivest-Shamir-Adleman [cryptosystem] |